# Characterization of micro pore optics for full-field X-ray fluorescence imaging

**S. An**[a,b]**, D. Krapohl**[a]**, B. Thörnberg**[a]**, R. Roudot**[c]**, E. Schyns**[c] **and B. Norlin**[a,1]

[a] *Mid Sweden University,*
  *Holmgatan 10, SE-851 70 Sundsvall, Sweden*

[b] *Lund University,*
  *MAX IV Laboratory, SE-221 00 Lund, Sweden*

[c] *Photonis France S.A.S.,*
  *Avenue Roger Roncier, 19100 Brive La Gaillarde, France*

E-mail: borje.norlin@miun.se

ABSTRACT: Elemental mapping images can be achieved through step scanning imaging using pinhole optics or micro pore optics (MPO), or alternatively by full-field X-ray fluorescence imaging (FF-XRF). X-ray optics for FF-XRF can be manufactured with different micro-channel geometries such as square, hexagonal or circular channels. Each optic geometry creates different imaging artefacts. Square-channel MPOs generate a high intensity central spot due to two reflections via orthogonal channel walls inside a single channel, which is the desirable part for image formation, and two perpendicular lines forming a cross due to reflections in one plane only.

Thus, we have studied the performance of a square-channel MPO in an FF-XRF imaging system. The setup consists of a commercially available MPO provided by Photonis and a Timepix3 readout chip with a silicon detector. Imaging of fluorescence from small metal particles has been used to obtain the point spread function (PSF) characteristics. The transmission through MPO channels and variation of the critical reflection angle are characterized by measurements of fluorescence from Copper and Titanium metal fragments. Since the critical angle of reflection is energy dependent, the cross-arm artefacts will affect the resolution differently for different fluorescence energies. It is possible to identify metal fragments due to the form of the PSF function. The PSF function can be further characterized using a Fourier transform to suppress diffuse background signals in the image.

KEYWORDS: X-ray fluorescence imaging; Micro pore optics (MPO); Point spread function (PSF); Hybrid pixel detector

---

[1] Corresponding author.

# Contents

## 1. Introduction

Micro X-ray fluorescence spectrometry (XRF) is a widely used non-destructive technique for performing elemental analysis down to the micrometre scale of length. Micro-XRF is capable of providing elemental distribution information in many fields, such as materials science, cultural heritage [1], [2] and planetary surface analysis [3], [4]. XRF imaging is typically performed using scanning or projection methods [5].

In the scanning method, XRF images are obtained by conducting a two-dimensional scan of a beam on the sample and collecting the fluorescence X-rays at every point of the map [6], [7], [8]. The scanning approach provides high chemical sensitivity and high spatial resolution with the help of a polycapillary focusing lens or synchrotron [9]. However, the spatial resolution of the obtained XRF image is limited by the spot size of the X-ray probe and scanning step size. It takes a significant amount of time to obtain elemental images for large areas measured with high spatial resolution.

Full-field X-ray fluorescence imaging (FF-XRF) is a projection method that allows for statically resolved X-ray spectroscopy [10], [11]. It can be used to map full sample areas with fair position and energy resolution. Primary X-rays are irradiated on a large area of the sample. The emitted XRF from the sample is collimated by X-ray optics (or a pinhole collimator [12], [13]) then guided to the pixel detectors. Several XRF imaging systems have been developed using X-ray optics. X-ray optics can be manufactured with different micro-channel geometries, such as square [14], hexagonal [15] or circular channels [16]. Although X-ray focusing optics have been significantly refined, they usually suffer from relatively low collection efficiency. In our previous research, a comparison study between a circular-channel polycapillary array and a pinhole collimator was conducted. No significant transmission intensity gain was observed when using a lead glass polycapillary array [17]. Moreover, circular-channel optics are not considered true focusing devices due to the fact that the circular channels are mirrors with very short focal lengths and, hence, rays reflecting from a single channel diverge. Square-channel micro pore optics (MPOs), sometimes referred to as 'square multi-channel plate optics', 'multi-pore optics' or 'lobster-eye optics' are an attractive option because of their efficiency, which results from the corner cube effect, when compared to other possible channel cross sections [14]. For square-channel MPOs, X-rays that reflect off the square pore sides form a central focus (odd number of reflections in one plane) or line focus (even number of reflections in one plane), giving a cross-arm point spread function (PSF). Ideally, one reflection each in two orthogonal planes will contribute to a focused central spot. The understanding of the square MPO properties relies on the modelling of ideal structures, as shown by one study carried out already in 1991 [14]. XRF imaging using a perfect planar square-channel MPO has been further investigated through the modelling of the PSF, whereby Monte Carlo simulations were developed to evaluate the effects of instrument geometry and MPO characteristics, including various types of defects [18]. However, the MPOs have imperfections that might add complexity to applications and create difficulties for understanding the performance of MPOs in practice.

In a collaborative study, we investigated the usage of a square-channel MPO in an FF-XRF imaging measurement system. The imaging system consists of a commercially available square-pore MPO and a Timepix3 readout chip with a silicon detector. Using this setup, the influence of geometrical parameters and X-ray energy on the PSF profile is shown. Transmission through MPO channels and variation of the critical reflection angle are characterized by the measurement of metal fragments with different fluorescence energies. The hypothesis is that the energy sensitivity of the MPO can provide useful energy information directly from artefacts in the image caused by the PSF profile. An alternative method for indirect energy measurement is to vary the angle between two MPOs mounted in series [19]. MPOs act as a filter for X-ray energy since the angle of total reflection for X-rays is energy dependent. Our purpose is to identify elemental imaging applications and spatial resolution limitations using MPOs.

## 2. Instrumentation

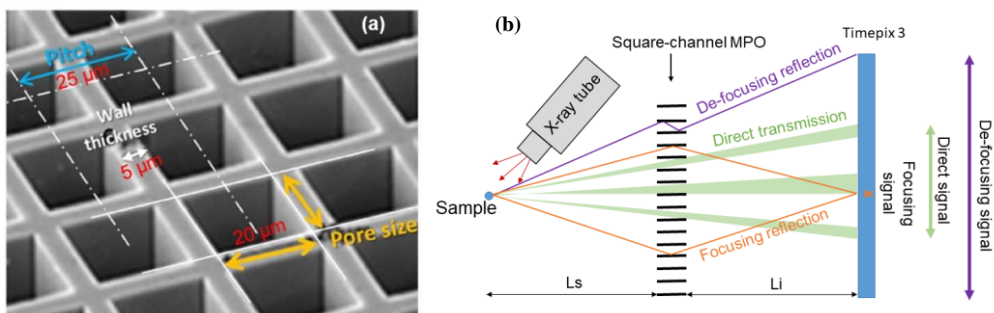

**Figure 1.** Setup: (a) SEM image of the square-channel MPO, (b) Sketch diagram of the XRF imaging setup with a square-channel MPO, containing different reflection components and their influence on the resolution.

The experimental setup of the FF-XRF imaging system using a flat square-channel MPO is shown in Figure 1. An X-ray tube with a silver target was used. The MPO was mounted at the midpoint between the sample and the pixel detector. Spectral resolved images of the fluorescence signal passing though the MPO are retrieved using a Timpix3 system. X-rays that enter the MPO at angles to a channel axis smaller than the critical angle will either pass through the channel or be submitted to one or a series of reflections with the channel walls. X-ray photons undergoing an odd number of reflections in one plane will be focused on the image plane, while direct transmission or double reflection will deteriorate the spatial resolution. With equal distances Ls and Li, between sample and MPO and between MPO and detector, the conditions of the image plane to obtain true focusing without magnification are fulfilled.

### 2.1 Timepix3 pixel detector

The Timepix3 readout chip is developed by the Medipix3 group at CERN. This ASIC is equipped with eight data channels that are data driven and zero suppressed making it suitable for particle tracking and spectral imaging [20]. The chip is designed in 130 nm CMOS and contains a 256 × 256-pixel matrix (55 × 55 μm$^2$). The Timepix3 can record time of arrival (ToA) and time over threshold (ToT) simultaneously for each pixel. As channel responses can never be identical, it is necessary to perform an energy calibration for each pixel. The calibration procedure is performed by taking measurements of X-ray fluorescence from five known elements: titanium, iron, copper, zirconium and silver plate. Moreover, a Timepix3 readout chip can be combined with different sensor materials, such as Si, GaAs or CdTe. A Timepix3 device with a 300 μm thick p-on-n Si detector was used in this study. After calibration, the energy resolution at 8.04 keV was 1.12 keV [17].

### 2.2 Micro pore optics plate

The MPO is a multichannel reflective X-ray lens of leaded glass composed of an array of microscopic square-section channels whose walls act as imaging mirrors, as seen in Figure 1 (a). X-rays that enter the optic at an angle smaller than the critical angle to a channel axis will either pass through the channel or undergo one or a series of reflections [1], [4], [19]. The MPO is shaped as a square with 20 mm sides and 1.2 mm thickness. The channel width is 20 μm square with 25 μm channel pitch giving an open area ratio of 60 %. The channel walls are coated with a 25 nm layer of iridium, a heavy reflecting material, to achieve a flat surface and hence improve optical properties [21]. The approximate critical angle ($\theta_c$) for total external reflection is given by the surface material atom number (Z), mass (A), density (ρ) and energy (E) of the X-rays using the following equation [22]:

$$\theta_c [deg.] \approx \frac{1.651}{E[keV]} \sqrt{\frac{Z}{A} \rho[g/cm^3]} \tag{1}$$

## 3. Results and discussion

### 3.1 Point spread function

In a realistic imaging system, each point of a measured image is blurred by the effects of neighbour points compared to an ideal image with signal in only one single pixel. That blurring is called a point spread function (PSF). A PSF is widely used to characterize the spatial resolution of imaging systems. The main features of a square-channel MPO are a central spot (which is desirable for image formation), two perpendicular lines forming a cross, and weaker intensity in the quadrants delimited by the cross. The central spot intensity, or focusing reflection in Figure 1, is the result of an odd number of reflections in one plane [18]. A second effect is from the diffuse signal from direct transmission. Moreover, an even number of reflections in one plane will contribute to the de-focusing reflection [23] and hence to the outer wings of the cross-arms.

   A flat-field image of a Cu plate was obtained and implemented to cancel the effects of image artefacts caused by variations in the pixel sensitivity of the detector. Measured PSFs after flat-field correction for different distances are shown in Figure 2. At a long working distance, the cross size is increased due to projection magnification (see Figure 2 (a)-(c)). However, when the MPO is placed non-symmetrically in the setup, the result is a blurred image, as shown in Figure 2 (d). This can be explained by the magnification effect due to non-true focusing.

To achieve a better spatial resolution, the Ls and Li distance should be as small as possible and the MPO located at the midpoint between the sample and pixel detector. However, a few centimetres of space need to remain for the X-ray tube tip to irradiate a large area of the sample. An alternative method to reduce the working distance is to tilt the sample table at 45° to the axis, as described in [24]. When tilting the sample stage, the XRF image is distorted on the image plane. Software can correct those distortions by warping the image with a reverse distortion.

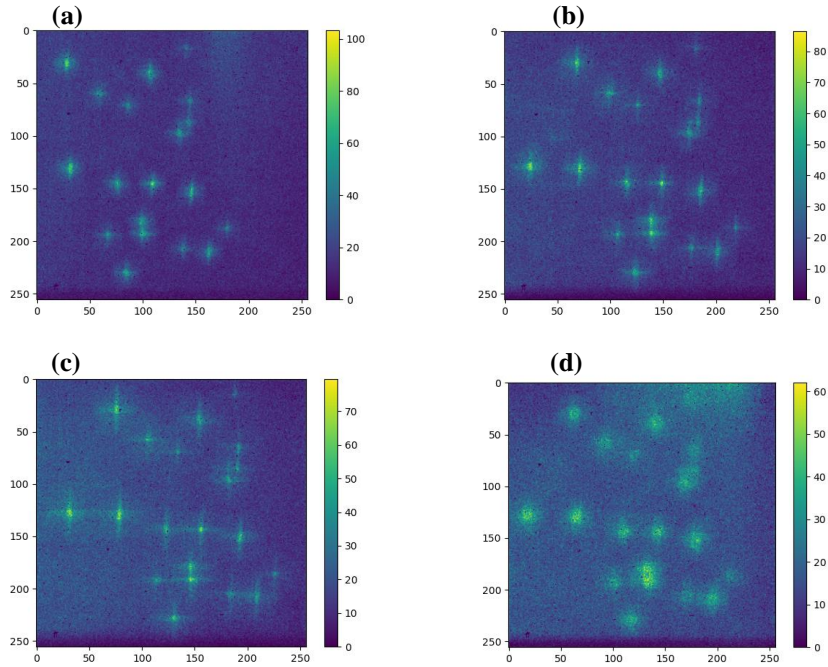

**Figure 2.** Measured PSF of MPO for Cu particles at different geometry arrangements: (a) Ls = Li = 2.5 cm; (b) Ls = Li = 3.5 cm; (c) Ls = Li = 4.5 cm; (d) Ls = 5cm, Li = 2 cm.

**3.2 Energy dependence**

Measurements with a point source are useful for showing the modulation of the PSF with the MPO position, but it is also worth knowing the PSF response for different energy levels with respect to fluorescence imaging. A sample with two point sources (a mixture of Ti and Cu particles) has been analysed. This sample was exposed at 13 kVp with 500 µA emission current for two hours. Depending on the fluorescence energy, we plot the image for the energy ranges 2.5 keV to 5.5 keV and 6 keV to 9 keV, thus an elemental map of the distribution of Ti and Cu is obtained, as shown in Figure 3 (b) and (c). These images reproduce the 'true' sample shown in the photo in Figure 3 (a).

An important feature of these PSFs is the presence of a cross centred on the main spot. These crosses have a detrimental effect on the image resolution. Thus, the intensity and reach of the cross-arms are worth characterizing. Both the object material and the system parameters influence these PSFs. The intensity of the cross-arms in the region around the PSF central spot depends on geometric factors, and reflectivity, which changes with the material and X-ray energy [18]. For a fixed geometry, the material and the X-ray energy will mainly affect the length of the cross-arms. To exemplify how PSF profiles look, we extracted the full 2.5 keV to 9.0 keV spectrum intensity of the horizontal arm and the vertical arm from one Ti particle and one Cu particle, and plotted the central spot profiles in Figure 3 (d) and (e). Each profile is achieved by averaging three nearby row or column profiles. The PSF centre peak is due to focusing by an odd number of reflections in one plane in the MPO walls. At low energy, the critical angle will be wider. Hence, a higher PSF centre peak is expected for Ti since the signal can be reflected by more MPO channels. The Kα emission line is 4.5 keV for Ti and 8.0 keV for Cu. The calculated critical angle is ~1.10° for Ti, and ~0.62° for Cu fluorescence, using the density of the iridium coating in Equation 1. Using the same argument, the Cu PSF centre peak intensity is expected to be lower than for Ti. As expected, the signal-to-noise ratio is worse for Cu.

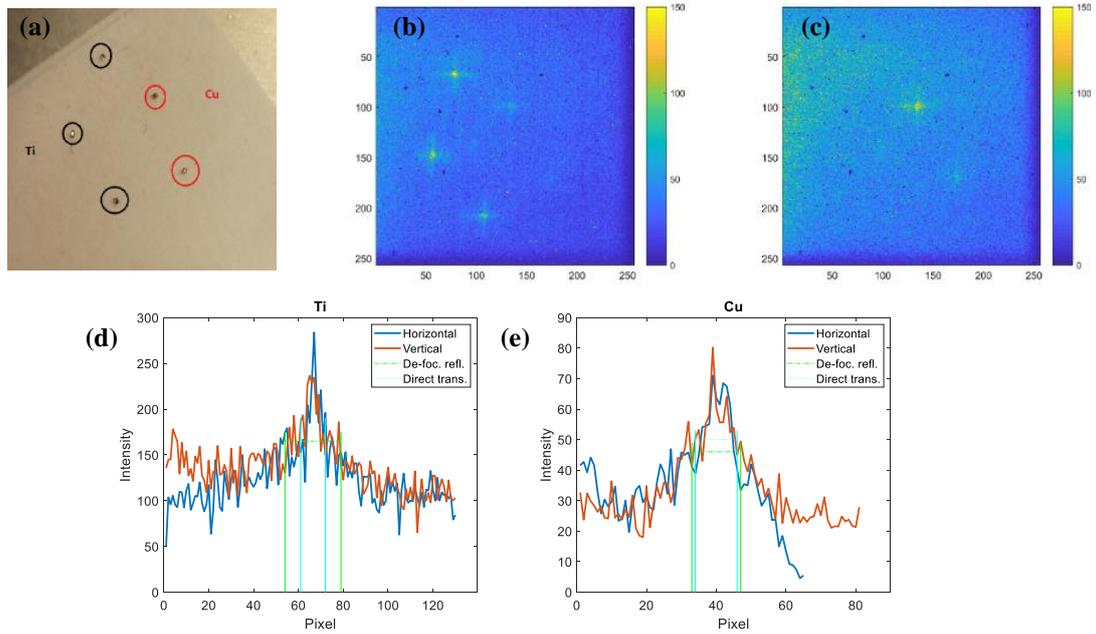

**Figure 3.** Example of PSF's from two materials resolved in one exposure image: (a) sample of metal particles; (b) PSF of Ti particles at 4.5 keV; (c) PSF of Cu particles at 8.0 keV; (d) central spot profiles of the MPO with Ti particle; (e) central spot profiles of the MPO with Cu particle. The expected FWHM due to components from two reflections and zero reflections are indicated in (d) and (e).

When it comes to arm length, wider spreading is expected for photons from Ti particles that experience double reflections in one plane in the channels, as illustrated in Figure 1, since the critical angle is greater for lower energy. In Figure 3 (d) and (e), the expected FWHM of photons undertaking de-focusing reflection is marked. The detector is also exposed to direct transmission beams with no interaction with the MPO walls. These direct transmissions through the optic is energy independent since it is limited only by the channel geometry. The expected FWHM for direct transmission is also marked to allow it to be compared with the measurement. In summary, the measurements shown in Figure 3 indicate that the increased arm length of Ti compared to Cu can possibly be used to distinguish these two elements in case a non-energy-resolving detector system were to be used. However, the identification is complicated by the fact that lower energy does not only increase the arm length but also increases the PSF centre intensity. Hence, the actual PSF profiles for different elements might be difficult to distinguish. Another problem is that the possibility to accurately measure the arm lengths will vary if the background level increases or decreases.

## 3.3 Fast Fourier transform

To suppress the diffuse background signal in the measured images, a fast Fourier transform (FFT) and its inverse were applied to the measurements. Previous works have applied a deconvolution algorithm [25] to a square-channel MPO image to reduce the impact from a non-ideal PSF and thus enhance the resolution [1]. Another motivation to further process the measurements is the expectation to see a pattern hidden in the diffuse background, along the arms of the PSF. A Gaussian-shaped window function was applied to individual PSFs of Cu from Figure 2 and of Ti from Figure 3, followed by an FFT for each PSF.

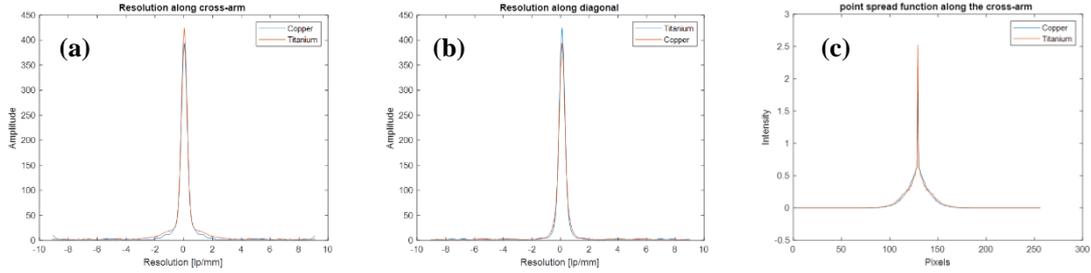

**Figure 4.** Amplitude transfer functions achieved from a fast Fourier transform of PSFs, along (a) and diagonal to (b) the cross-arms. PSFs (c) achieved from inverse Fourier transform average of three PSFs, where background is strongly suppressed.

It should be noted that the phase information from the FFT output is neglected such that the spatial position of the impulse response is suppressed. Hence, the achieved amplitude transfer functions from known metal fragments can be averaged. Figure 4 shows FFT results for Cu and Ti, averaged from three PSFs. Figure 4 (a) shows that along the cross-arm there is a difference in the shape, indicating that Ti has higher resolution. Along the diagonal between cross-arms there is no significant difference between the two materials seen in Figure **4** (b). After applying inverse Fourier transform to the achieved amplitude transfer functions, idealised PSFs with supressed background will be achieved. The achieved resolution is slightly below 1 lp/mm. The relatively high background level seen in Figure 3 can be reduced to close to zero in Figure 4 (c). In this study, it could not be verified if an oscillation pattern occurred along the cross arms, although some vague indication to a repetitive pattern is seen in the amplitude transfer function for Cu.

## 3.4 Discussion

Although the features of the PSF have a negative impact on the XRF image when using a square-channel MPO, we have demonstrated that the cross-arm length of the PSF is energy dependent, using fluorescence from Ti and Cu. This feature is similar to a wavelength dispersive XRF spectrometer. The different energies of the characteristic radiation emitted from the sample are reflected in different directions and locations by the MPO walls. By measuring the arm length, we might be able to calculate the photon energy even when using non-energy-resolving detectors. One problem is that direct transmission and de-focusing reflection are merged with the peak signal due to focusing reflection. The direct transmission is determined by energy-independent MPO parameters, such as thickness, pore size and channel pitch; this will create a diffuse background patch and affect the determination of the de-focusing reflection arm length.

To improve the imaging performance of an FF-XRF imaging system, the MPO parameters can be optimized during manufacturing for a dedicated application. Lens thickness, channel size, and internal coating are the main parameters that can be adjusted in the MPO design. In previous ray tracing studies, two different methods have been proposed to reduce the impact on the MPO [18]. The first approach involves an array of square pores with a random orientation of the square cross section. This has been simulated for large distances (10 cm) and short distances (0.5 cm).

The image obtained using the randomly oriented squares is not very different at a short distance to the image obtained using a regular MPO [18]. This can be explained by the limited number of channels involved for each point of the extended source, which is also the origin of the periodicity of the PSF discussed above in the case of a regular MPO. However, this presents practical difficulties when manufacturing the MPO. The second approach involves the precise rotation of a regular MPO to negate the effect of the PSF, which provides good results at a short distance. However, distances down to 0.5 cm are a challenge for the mounting setup. Our setup needs at least ~3 cm of space for the X-ray tube tip between the sample and collimator for the large irradiation area of the sample.

To further characterize the channel transmission and energy dependence of total reflection of the square MPO, post-processing using a Fourier transform has been applied. The relatively high diffuse background signal from the original measurements can be reduced to close to zero using this methodology. It is possible to reduce the background due to averaging several PSFs. Different PSFs can vary, however, due to fragment size and orientation; we might not measure true the PSF for all fragments. This can explain why, in Figure 4 (c), the difference between Ti and Cu PSFs is supressed after averaging.

## 4. Conclusion

The purpose of the current study was to identify capabilities and limitations using MPOs. A full-field XRF camera setup with a square-channel MPO and Timepix3 readout chip was characterized in this study. The instrument was validated against theory using X-ray fluorescence imaging experiments with varying parameters for X-ray fluorescence energy, distances and readout energy discrimination. We have investigated the influence of these parameters on the intensity and spatial resolution of an X-ray fluorescence imaging experiment. One interesting feature of the single square-channel MPO setup is the ability to obtain spatially resolved images, where the arm length in the PSF indicates energy even if the imaging sensor does not resolve energy. Post-processing using a Fourier transform is a promising method for background suppression if the PSFs are generated from small enough fragments.

## Acknowledgements

The authors acknowledge funding from the Swedish Knowledge Foundation (ImSpec-Multiple energy band imaging spectroscopy for material and object classification). The measurements were performed using detectors, software and a readout system developed by the MEDIPIX collaboration hosted by CERN.